\documentclass{article}
\usepackage{arxiv}

\newif\ifuseclearpage
\useclearpagetrue
\newif\ifuseauthors
\useauthorsfalse

\usepackage[utf8]{inputenc} 
\usepackage[T1]{fontenc}    
\usepackage{hyperref}       
\usepackage{url}            
\usepackage{booktabs}       
\usepackage{amsfonts}       
\usepackage{nicefrac}       
\usepackage{microtype}      
\usepackage{lipsum}		
\usepackage{graphicx}

\usepackage{array}
\usepackage{amssymb, amsmath, amsthm}
\usepackage{graphicx}
\usepackage{lmodern,url}
\usepackage{makecell} 
\usepackage{cancel}
\usepackage{multirow}
\usepackage{microtype}
\usepackage{lscape}
\usepackage{xspace}
\usepackage{xcolor}
\usepackage{siunitx}
\usepackage{todonotes}
\usepackage{booktabs}
\usepackage[ruled,vlined]{algorithm2e}
\usepackage{optidef}
\usepackage{mathdots}
\usepackage{subcaption}
\usepackage{csquotes}
\usepackage{caption}
\usepackage{verbatim}
\usepackage{ulem}
\usepackage{bibunits}
\graphicspath{{Figures/}}

\usepackage{amsmath}
\usepackage{tikz}
\usetikzlibrary{fit}
\usetikzlibrary{positioning}
\tikzset{%
  highlight/.style={rectangle,rounded corners,fill=red!15,draw,fill opacity=0.25,thick,inner sep=0pt}
}

\allowdisplaybreaks
\title{Mechanics of Pandemics}

\usepackage{authblk}

\author[a,b$\dagger$*]{Seba Contreras}
\author[a,b$\dagger$]{Philipp D\"onges}
\author[a,b$\dagger$]{Laura M\"uller}
\author[a,b$\dagger$]{Piklu Mallick}

\author[c]{Sydney Paltra}
\author[d,e,a]{Ulrik Hvid}
\author[f,g,h]{Robyn J. N. Kettlitz}

\author[i]{Andreas Reitenbach}
\author[a,b]{Rodrigo Amaral Lind}

\author[j,k]{Maíra Aguiar}
\author[l]{Philip Bechtle}    
\author[m]{André Calero Valdez}
\author[a]{Ronja Gronemeyer} 
\author[f,h]{Manuela Harries}   
\author[n,o]{Veronika K. Jaeger}
\author[n,o]{André Karch}
\author[f,h]{Carolina J. Klett-Tammen}  
\author[p,q,r]{Peter Klimek}
\author[s,t,m]{Mirjam E. Kretzschmar}
\author[c]{Kai Nagel}
\author[d,e]{Bjarke Frost Nielsen}
\author[u]{Barbara Prainsack}
\author[u]{Isabella M. Radhuber}
\author[d,v]{Lone Simonsen}
\author[d,e]{Kim Sneppen}
\author[n,o]{Janik Suer}

\author[a,b*]{Viola Priesemann}

\affil[a]{Max Planck Institute for Dynamics and Self-Organization, G\"ottingen, Germany.}
\affil[b]{Institute for the Dynamics of Complex Systems, University of G\"ottingen, G\"ottingen, Germany.}
\affil[c]{Chair of Transport Systems Planning and Transport Telematics, Technische Universität Berlin, Berlin, Germany}
\affil[d]{PandemiX Interdisciplinary Center for Pandemic Signatures, Copenhagen, Denmark} 
\affil[e]{Biocomplexity, Niels Bohr Institute, University of Copenhagen, Copenhagen, Denmark} 
\affil[f]{Department for Epidemiology, Helmholtz Centre for Infection Research, Braunschweig, Germany}
\affil[g]{PhD Programme “Epidemiology” Braunschweig-Hannover, Germany}
\affil[h]{German Centre for Infection Research (DZIF), partner site Hannover-Braunschweig, Braunschweig, Germany}
\affil[i]{Karlsruhe Institute of Technology, Karlsruhe, Germany}
\affil[j]{BCAM - Basque Center for Applied Mathematics, Bilbao, Spain}
\affil[k]{Ikerbasque, Basque Foundation for Science, Bilbao, Spain}
\affil[l]{Physikalisches Institut, Universität Bonn, Bonn, Germany}
\affil[m]{Institute of Multimedia and Interactive Systems, University of Lübeck, Lübeck, Germany}
\affil[n]{Institute of Epidemiology and Social Medicine, University of Münster, Münster, Germany}
\affil[o]{Interdisciplinary Center for the Mathematical Modeling of Infectious Disease Dynamics (IMMIDD), University of Münster, Münster, Germany.}
\affil[p]{Institute of the Science of Complex Systems, CeDAS, Medical University of Vienna, Vienna, Austria} 
\affil[q]{Complexity Science Hub Vienna, Vienna, Austria} 
\affil[r]{Supply Chain Intelligence Institute Austria, Vienna, Austria} 
\affil[s]{University Medical Center Utrecht, Utrecht University, Utrecht, The Netherlands.}
\affil[t]{Center for Complex Systems Studies (CCSS), Utrecht University, Utrecht, The Netherlands}
\affil[u]{Department of Political Science, University of Vienna, Vienna, Austria}
\affil[v]{Department of Science and Environment, Roskilde University, Roskilde, Denmark} 

\affil[ ]{* Corresponding Author: Viola Priesemann (viola.priesemann@ds.mpg.de), Seba Contreras (seba.contreras@ds.mpg.de)}
\affil[ ]{{$\dagger$} These authors contributed equally}

\renewcommand{\headeright}{ }
\renewcommand{\undertitle}{ }

\begin{document}
\maketitle
\clearpage
\begin{abstract}
COVID-19 and previous pandemics have shown how diseases can disrupt, threaten, and transform daily life. Since pathogens and societies are continuously evolving, every pandemic is different. However, certain fundamental principles of disease transmission appear to hold true across different outbreaks. These ``mechanisms'' are grounded in natural laws or the very structure of our biology and societies. This paper compiles ten fundamental mechanisms, curated by a multidisciplinary team with backgrounds spanning public health, medicine, epidemiology, political science, mathematics, physics, and psychology. These mechanisms, although perhaps underappreciated, substantially shape how pandemics unfold and are controlled. The better we succeed in understanding these mechanisms and establishing this knowledge in our societies, the better we will be able to prepare for future pandemics and respond appropriately when they occur.  
\end{abstract}


\begin{bibunit}
\clearpage
\section*{Introduction}

In our ever-changing world, pandemics will continue to pose a substantial risk to the well-being of our societies. While we cannot predict when or how the next crisis will emerge, its arrival appears inevitable because pathogens and societies coevolve: emerging infectious diseases drive societies to adapt through new technology, medicine, and preventive measures, thereby applying developmental pressure and prompting pathogen selection and ongoing evolution ~\cite{nielsen2022lockdowns,morens2009pandemic, brett2025collateral,wasik2019onward,lloyd2005superspreading}. As a result, no two pandemics are identical. Each pandemic possesses unique characteristics defined by, e.g., its transmission routes, clinical manifestations, and societal impacts~\cite{morens2009pandemic,brett2025collateral,medley2017emerging}. Yet, whether hidden deep down or visible in plain sight, fundamental similarities exist in how outbreaks unfold and can be controlled. These similarities arise from recurring epidemiological, behavioral, and societal dynamics that shape the spread of infectious diseases across contexts. Because these ``mechanisms'' are grounded in natural laws and the structure of our biology and society, they are likely to remain relevant in future pandemics even when the pathogen itself is novel. Importantly, these mechanisms reflect fundamental principles that can inform public health decision-making even before detailed quantitative evidence becomes available. Because they capture fundamental properties of disease spread and societal response, they remain valuable under substantial uncertainty, including when the characteristics of a novel pathogen are still only partially understood. In this paper, we therefore compile and synthesize core mechanisms of pandemics across disciplines, many of which are familiar to specialists but less accessible to decision-makers outside those fields, and present them in a form accessible to readers involved in pandemic preparedness and decision-making, including policymakers, public health practitioners, and interdisciplinary researchers. We refer to them as the mechanics of pandemics.

Whether an outbreak escalates into a pandemic depends on the characteristics of the pathogen and the society in which it spreads, encompassing behavior, governance, and technological capabilities~\cite{fraser2004factors,kretzschmar2020impact,morens2020emerging}. Although definitions vary across the literature, here we define a pandemic as an epidemic that has spread across multiple countries or continents, typically affecting a large proportion of the population~\cite{porta2014dictionary,morens2009pandemic}. Based on lessons from recent pandemics, some fundamental epidemiological and transmission principles have become widely recognized, such as the initial exponential growth of cases. 
At the same time, outbreaks are shaped by dynamic social and behavioral responses, such as changing risk perception and solidarity circles. These mechanisms and others will be discussed in this paper. While they can be considered individually, pandemics emerge from their interaction, and we therefore also discuss how these mechanisms reinforce or counteract one another in practice.

To obtain a broader perspective on the mechanisms that will shape our understanding of the next pandemic, we invited experts who work at the interface of epidemiology, sociology, physics, and public health to each distill a specific mechanism underlying disease transmission, control, or prevention.  
The contributions were iteratively reviewed, edited, and discussed (see Appendix~\ref{SI:contributions}). Rather than advancing a single disciplinary perspective, this paper aims to integrate complementary insights from multiple fields into a coherent conceptual synthesis. Importantly, the work is not intended as a unified predictive model; instead, the mechanisms are presented as conceptual principles rather than empirically calibrated predictions. Our examples are not intended to be exhaustive; rather, they synthesize expert perspectives and lessons learned from recent pandemics, focusing on the fundamental aspects that can guide our understanding and mitigation of future pandemics.

\ifuseclearpage\clearpage\fi
\section{Effective prevention requires monitoring the system rather than the outbreak}
\ifuseauthors\textit{Maíra Aguiar}\fi
\label{mec:onehealth}

Emerging and re-emerging infectious diseases are not isolated events but natural outcomes of evolutionary and ecological dynamics \cite{saldana2024modeling,pauciullo2024spillover}. Pathogens continually evolve in animal reservoirs, environmental niches, and human populations, shaped by selective pressures including immunity, interventions, and habitat change. The critical transition occurs at spillover: when a pathogen crosses from its reservoir, animal, plant, or abiotic, into a susceptible human host \cite{plowright2017pathways,morens2020emerging}. When introduced into a largely immunologically naive population, these pathogens may trigger extensive outbreaks either due to their natural transmissibility or stochastic amplification, yet most spillover events fail to sustain onward transmission \cite{borremans2019cross}.

It is essential to distinguish the initial spark of spillover from the subsequent establishment of sustained human-to-human transmission. Each spillover represents an ecological accident -- for example, the alignment of a shedding animal and a susceptible human -- that has the potential to ignite epidemic spread \cite{lloyd2009epidemic}. However, the magnitude and public-health impact of an outbreak are not determined by transmission potential alone, but also by population vulnerability, heterogeneity in disease severity, and condition-specific mortality risk \cite{aguiar2020condition}. Near the epidemic threshold, random fluctuations in contact patterns, importation timing, vector abundance, or environmental conditions can generate large but transient outbreaks even when transmission is subcritical or only marginally supercritical, or extinguish sparks with pandemic potential by chance \cite{lloyd2005superspreading,pisaneschi2025fewmosquitoes}. When a pathogen with epidemic potential ($R_0>1$) continues to spread in a population, new cases are expected to grow exponentially, and stochastic extinction soon becomes unlikely. This mechanism therefore interacts directly with early exponential growth: once sustained transmission is established, even short delays in detection or response can translate into large increases in cumulative incidence. It also interacts with superspreading, since rare spillover events or small transmission clusters may either disappear unnoticed or expand rapidly before routine case-based surveillance detects a clear signal. Prevention based only on observed outbreaks therefore responds to a late manifestation of a deeper system-level transition.

Therefore, when an outbreak does occur, reactive interventions---introduced only after detection---are, by design, deployed too late and are consequently costly \cite{contreras2021low,wasik2019onward,aguiar2021critical,muller2025optimizing}. Effective prevention should also act upstream, by detecting instability within the broader One Health system before it becomes visible as a human outbreak. Preparedness, then, rests on continuous, multi-scale surveillance of a shared biological system, defined by interdependent species and environments \cite{grange2021ranking,pauciullo2024spillover}. The applicability of this mechanism depends on surveillance capacity, data integration, and institutional trust. In high-resource settings, system monitoring may include wastewater sampling, genomic surveillance, sentinel animal monitoring, and environmental data streams; in lower-resource settings, it may rely more heavily on community-based surveillance, syndromic reporting, animal-health networks, and targeted ecological monitoring. Because upstream warning signals are often indirect and probabilistic, effective use also requires careful risk communication, so that authorities can justify early action before visible human outbreaks occur without undermining trust through poorly explained false alarms. Recognizing this interdependence is a scientific and societal imperative: preventing the next pandemic begins with acknowledging that our vulnerabilities are, and will remain, collective \cite{saldana2024modelling,morens2020emerging,wasik2019onward}. Further implications of this system-level view are discussed in Appendix~\ref{SI:system_monitoring}.

\begin{figure}[ht!]
    \centering
    \includegraphics{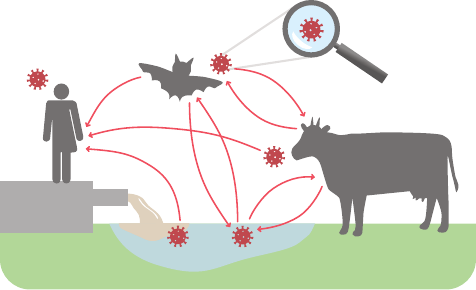}
    \caption{Pathogens evolving in environmental reservoirs can spill over to the human population, each spillover being a ``spark'' with the potential to ignite epidemic spread within the immunologically naive population.
    These sparks unfold against slower-moving structural variables, including climate, land use, and mobility, which determine the ultimate outcome, ranging from transmission dies out to escalation into a global outbreak.
    }
    \label{fig:placeholder}
\end{figure}

\ifuseclearpage\clearpage\fi
\section{The effect of combined NPIs can be redundant}
\ifuseauthors\textit{Peter Klimek}\fi
\label{mec:redundant_npis}

In the early phase of a pandemic, when effective vaccines and medications are unavailable, non-pharmaceutical interventions (NPIs), such as mobility restrictions and mask mandates, constitute the primary means of mitigating pathogen transmission. Their effectiveness is measured by their reduction of the effective reproduction number, and several studies have quantified their impact in different settings \cite{brauner2021inferring,haug2020ranking}. Although effective, NPIs entail substantial societal costs; therefore, identifying the most cost-effective and least invasive combinations is crucial for both compliance and effectiveness.

However, because NPIs may share overlapping mechanisms of action, the effect of combined NPIs is often redundant, which means that the combined effect of two measures is smaller than the sum of their individual effects. 
For example, school closures would incentivize parents to work remotely, thereby reducing their mobility so that additional ``work from home'' mandates would not affect them. However, measures can also be synergistic and thus more effective when combined, e.g., testing and targeted isolation \cite{contreras2021challenges,muller2025testing}, or paid sick leave and quarantine adherence \cite{radhuber2025inclusive,andersen2023does}. This systemic perspective can be formalized as an ``algebra of NPIs'', where interventions interact through context-dependent rules that capture redundancy, synergy, and diminishing returns (see Appendix~\ref{SI:redundant_npis}).

All of this has direct consequences for policy design. First, it highlights that policymakers should not rely solely on static rankings of interventions, but instead assess how measures interact within the existing portfolio of controls \cite{haug2020ranking}, though estimating these interactions in practice can present substantial challenges (see Appendix~\ref{SI:challenges_redundant_npis}). Second, if interventions have redundant effects, then stronger measures are over-proportionally difficult to achieve. The need for such strong measures can often be avoided through early, decisive action. Third, entirely different combinations of NPIs can achieve the same epidemiological effect \cite{lasser2022assessing,brauner2021inferring}. In summary, NPIs do not operate as independent levers but as interacting elements of a complex system.
Policies should therefore be designed and evaluated not only in terms of their direct epidemiological outcomes but also in terms of their broader societal impacts and feasibility in real-world governance \cite{haug2020ranking,glaubitz2024social,petherick2021worldwide,muller2025optimizing}.

\begin{figure}[ht!]
    \centering
    \includegraphics{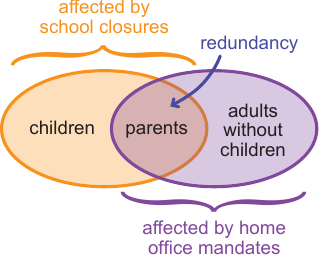}
    \caption{Non-pharmaceutical interventions often have redundant or synergistic effects when combined. One plausible explanation for this is the overlapping effect they may have on contact patterns and transmission probabilities.}
    \label{fig:redundant_npis}
\end{figure}

\ifuseclearpage\clearpage\fi
\section{Reducing event size can reduce new infections quadratically}
\ifuseauthors\textit{Kai Nagel and Sydney Paltra}\fi
\label{mec:quadratic_reduction}

Reducing the attendance at social gatherings (i.e., reducing the density of people at a given event) reduces disease transmission. However, the mathematical relationship is, in many cases, stronger than commonly assumed: the expected number of new infections decreases quadratically with the reduction in attendance \cite{paltra2024effect}. For example, reducing attendance to half reduces infections to one-quarter of the original level. The mechanism is as follows: Restricting the maximum attendance at events decreases both the number of infections that can be brought in and the number of susceptible individuals who can become infected. This results in a multiplicative (i.e., quadratic) reduction in the expected number of new infections: If only a fraction $\alpha < 1$ of a base turnout attends, then only $\alpha$ can cause infection, and only $\alpha$ can become infected. As a consequence, expected new infections scale as $\alpha^2$ \cite{muller2021predicting, paltra2024effect,Vecherin2022Assessment}. 

When wanting to extend this quadratic effect from a single gathering to all social gatherings, it implies that evenly thinning out attendance across activities is more effective than fully shutting down some while leaving others untouched. For example, assume there are only two independent activities, A and B, with the same transmission-relevant characteristics: contact intensity, room size, indoor/outdoor activity, and duration. Reducing attendance at both activities to half yields an overall activity reduction of half, but a reduction in expected infections to one-quarter. On the other hand, completely shutting down activity~B, but allowing individuals to pursue activity A as before, reduces the overall activity level again to half, but only reduces the expected number of infections to half (see Appendix~\ref{SI:quadratic_reduction}).

The effect has been shown for COVID-19 and also applies to other infectious diseases where the force of infection depends on the density of infectious individuals, i.e., when the well-mixed assumption holds. In such cases, preparedness should emphasize moderate restrictions across many sectors rather than complete closure of a few.

\begin{figure}[ht!]
    \centering
    \includegraphics{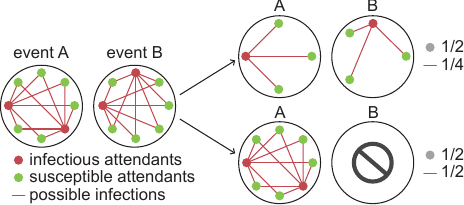}
    \caption{Restricting the maximum attendance at events decreases both the number of infections that can be brought in and the number of susceptible individuals who can become infected. This results in a multiplicative (i.e., quadratic) reduction in the expected number of new infections. As a consequence, even though in both scenarios above total attendance is halved, in the upper scenario the possible infections are reduced to 6, whereas in the lower scenario they are only reduced to 12.
    }
    \label{fig:quadratic_reduction}
\end{figure}

\ifuseclearpage\clearpage\fi
\section{Limiting sporadic contacts prevents infections between clusters}
\ifuseauthors\textit{Veronika K. Jaeger, Janik Suer, and André Karch}\fi
\label{mec:repeated_contacts}

Disease transmission dynamics depend not only on the number of contacts but also on their intensity. In this respect, human contacts can be broadly classified as repeated or sporadic, recognizing that real-world interactions span a continuum of contact frequencies and durations between these two extremes. Repeated contacts, such as those among family or co-workers, define clusters that limit the local outbreak sizes of respiratory infections. On the other hand, people have sporadic, one-time interactions, which allow pathogens to be transported between clusters. Understanding the interplay between these types of interactions is crucial for designing effective interventions.

In the context of respiratory infections, the number of individuals an infectious person can infect through repeated contacts, such as the daily interactions with the same household members and co-workers, is limited. Typically, such repeated contacts occur with a higher contact intensity and longer duration, leading to a high transmission probability per contact. In contrast, a set of daily-varying contacts, such as interactions on public transport or with strangers at work, provides a much larger set of potential transmission pathways, even if daily contact numbers are the same. Here, the transmission probability per contact is often smaller than for repeated contacts. Still, as the infectious period typically lasts for multiple days, individuals with varying daily contacts may infect a much larger pool of people. The difference in number of infected individuals between repeated and daily varying contacts becomes especially pronounced for pathogens with high transmissibility. In this case the susceptibles in the repeated contact are rapidly depleted while daily varying contacts result in a new pool of potentially susceptible individuals each day. 

Real-world contact networks are not tree-like but exhibit clustering: households, classrooms, and friend groups show high levels of triadic closures, the common “a friend of a friend is my friend” effect. This clustering further reduces epidemic spread, as many infection pathways overlap. For example, in a small household, the first infected individual can spread the pathogen to all other members. In that case, the second generation of infected cannot transmit the pathogen anymore as the pool of susceptibles has been depleted \cite{doenges2024sir}. Limiting social contacts to highly clustered groups, i.e., allowing individuals to only (or mostly) interact with their "social bubbles" of repeated contacts in case of an epidemic, can thus substantially reduce the effective reproduction number and the social burden caused by interventions at the individual level \cite{Danon2021HouseholdBubbles,Leng2021Effectiveness}.

\begin{figure}[ht!]
    \centering
    \includegraphics{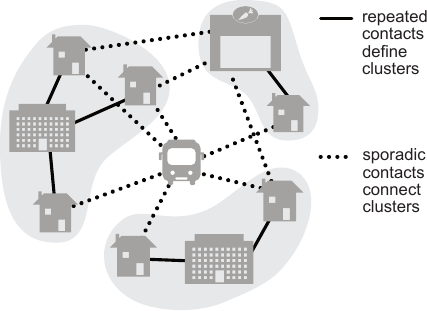}
    \caption{Contact networks are a combination of repeated and sporadic contacts. Repeated contacts, typically organized in clusters (e.g., workplaces or households), naturally limit the number of individuals an infectious person can infect. Sporadic contacts provide paths for a pathogen to spread from one cluster to another. Thus, limiting sporadic instead of repeated contacts is often better for epidemic control.}
    \label{fig:repeated_contacts}
\end{figure}

\ifuseclearpage\clearpage\fi
\section{Early and decisive action limits exponential pandemic growth}
\ifuseauthors\textit{...}\fi
\label{mec:exponential_growth}

At the onset of a pandemic, disease spread is facilitated by the lack of widespread immunity. Under such conditions, where most of the population is susceptible, classical mathematical models predict exponential dynamics. This leads to a simple exponential behavior, controlled by the average number of new cases generated by each infected individual. If the effective reproduction number $R$  is larger than one, $R > 1$, the number of infections grows exponentially. If, on the other hand, $R < 1$, the number of infections decrease exponentially over time (see Appendix \ref{SI:exponential_growth}). This has profound consequences for the early pace of an outbreak and for the design of intervention policies.

The first consequence concerns the timing of interventions: early action in the face of exponential growth means that delayed interventions must mitigate an exponentially larger incidence \cite{kompas2021health, balmford2020cross, muller2025optimizing}. An intervention deployed one generation later must control an outbreak that is about $R$ times larger, and after $k$ missed generations, the number of infections to be managed can grow by roughly $R^k$. This exponential growth means that a disease with $R > 1$ can escalate from a handful of cases to a major outbreak within a few weeks (depending on generation time). The second consequence concerns the intensity of interventions: mitigation efforts need to bring the reproduction number below the critical value of $R=1$. Otherwise, case numbers will continue to grow exponentially, 
although at a slower growth rate, meaning that the infection wave is mainly shifted in time rather than avoided \cite{dehning2020inferring}. 

In summary, early, sufficient interventions can stop exponential growth before case numbers become large. The window of opportunity in which a targeted effort can avoid a larger outbreak narrows quickly as cases accumulate. 

\begin{figure}[ht!]
    \centering
    \includegraphics{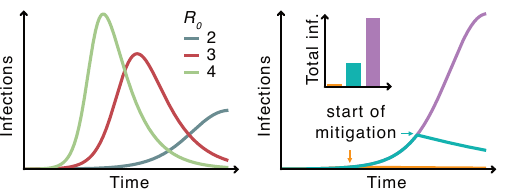}
    \caption{The mostly susceptible population at the onset of a pandemic leads to exponential growth of infections. During this time, each infectious individual infects on average $R_0$ other individuals over the course of their infectious period. As a result, even small delays in implementing control measures can lead to substantially larger outbreak sizes.}
    \label{fig:friendship_paradox}
\end{figure}

\ifuseclearpage\clearpage\fi
\section{Superspreading diseases are highly susceptible to restrictions on non-repeated contacts}
\ifuseauthors\textit{Ulrik Hvid, Lone Simonsen, Kim Sneppen, and Bjarke Frost Nielsen}\fi
\label{mec:superspreader}

Disease spread in real-world settings is highly heterogeneous. The term ``superspreading'' describes diseases with exceptionally high heterogeneity, in which a small number of individuals account for a large proportion of secondary cases \cite{lloyd2005superspreading,Nielsen2023Superspreading,Sneppen2021,althouse2020superspreading}. Such variability is often rooted in behavioral differences, e.g., some people having very many contacts, and others have few \cite{mossong2008social,Boily2009_HeterosexualRiskHIV1, Hvid2025}, but can also arise from biological processes, e.g., marked differences in viral shedding \cite{puhach2023sars}. We can categorize superspreaders based on whether superspreading is driven primarily by contact behavior or by infectiousness. The mechanism behind superspreading has direct consequences for disease spread and mitigation.

If superspreading is driven by individuals with many contacts, it is straightforward that interventions that limit the total number of unique contacts will prevent most of the cases they could generate. More interestingly, these superspreaders are also ``super receivers'', i.e., more likely to encounter the pathogen and be infected. As they are infected and immunized earlier in an outbreak, this lowers the threshold for herd immunity compared to homogeneous diseases \cite{Britton2020,Tkachenko2021Transient}. Moreover, potential superspreaders of this kind could be identified before infection, enabling highly effective targeted interventions like vaccination \cite{Hvid2025,Guzzetta2024mpox,Petersen2025_mpox_scenarios}.

\begin{figure}[ht!]
    \centering
    \includegraphics{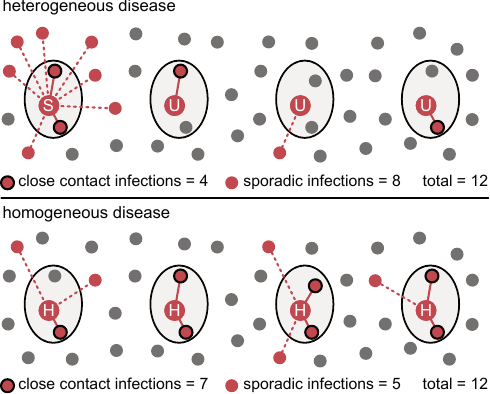}
    \caption{In diseases characterized by superspreading, a small subset of infected individuals (S) accounts for a disproportionate number of transmissions, often through sporadic, non-repeated contacts outside their regular contact network, while most individuals infect very few (U). Consequently, restricting non-repeated contacts can substantially reduce transmission. In contrast, for homogeneous diseases, individuals infect approximately the same number of people on average (H), so limiting sporadic contacts has a much smaller effect on overall spread.
    }
    \label{fig:superspreaders}
\end{figure}

Conversely, superspreaders of the second kind (higher infectiousness, e.g., some people expressing very high viral load~\cite{puhach2023sars}) are not necessarily ``super receivers'' and are thus not immunized earlier in an outbreak. Furthermore, they will infect not only a large fraction of their close contacts, but also of their sporadic, non-repeated contacts (see Fig.~\ref{fig:superspreaders}). Therefore, interventions limiting the number of non-repeated contacts will have a greater impact on transmission and containment than for a homogeneous disease with the same potential for spread \cite{Sneppen2021}. 

Ultimately, while superspreading appears to be a major asset to the pathogen, it is actually an Achilles’ heel. Identifying its nature early---via contact tracing and phylogenetics---enables interventions targeting specific transmission dynamics, effectively mitigating the disease's strongest pathways of spread \cite{leung2021transmissibility,Nielsen2023Superspreading,althouse2020superspreading}.

\ifuseclearpage\clearpage\fi
\section{Your friends are not ill if pandemic control is successful}\label{mec:friends_not_sick}
\ifuseauthors\textit{Viola Priesemann, Ronja Gronemeyer, Philipp Dönges and Piklu Mallick}\fi
\label{mec:friends_not_sick}

In successful pandemic mitigation, we face a strong discrepancy between subjective experiences and objective threats to public health. This discrepancy arises in a pandemic situation with (i) a high risk of severe disease requiring intensive healthcare, 
(ii) limited treatment capacity of the healthcare system, and (iii) effective interventions that avoid its collapse.

In more detail, for a pathogen with high severity, we see a very low prevalence threshold at which the health care system would be overwhelmed. Consequently, interventions must suppress transmission very early and keep it below this threshold to successfully mitigate morbidity and mortality. At such low prevalence, the probability that any given individual knows about an active case within their immediate social circle is negligible. Thus, due to the effective pandemic control, most individuals are unlikely to personally know anyone who is sick.

This phenomenon can be illustrated by comparing COVID-19 to a common cold. 
Early COVID-19 variants were associated with high hospitalization rates per infection (requiring intensive care). In regions with successful mitigation strategies, infection rates remained low, and few individuals knew someone who had contracted COVID-19; only a fraction of those cases were severe. In other regions with late or insufficient mitigation, many people knew of infections or even severe cases firsthand. In contrast, the common cold is considered a mild disease that does not require mitigation; consequently, most individuals know someone who has recently had a cold. 

In summary, successful pandemic control poses a communication challenge due to the \textit{prevention paradox}: the absence of severe illness in one's own social circle can be mistaken for overreaction rather than a sign of success \cite{messinger2020paradox}. Recognizing that not knowing anyone who is ill during a pandemic is a strong indicator of successful pandemic control is crucial for public understanding and support, and further implications of this prevention paradox are discussed in Appendix~\ref{SI:friends_not_sick}.

\begin{figure}[h!]
    \centering
    \includegraphics{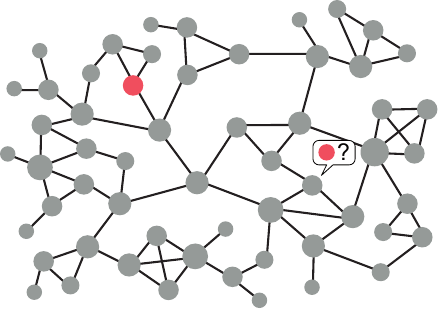}
    \caption{In the absence of a cure or vaccine, mitigation measures are necessary to prevent hospitals from being overburdened during a pandemic. Protecting hospital capacity when hospitalization rates are high implies keeping prevalence very low, so most people are unlikely to experience firsthand the pandemic's severity or personally know someone who is ill.}
    \label{fig:friends_not_sick}
\end{figure}

\ifuseclearpage\clearpage\fi
\section{Desynchronization of waves enables resource sharing}
\ifuseauthors\textit{André Calero Valdez}\fi
\label{mec:desynchronized_waves}

Epidemic growth from a networked perspective is a combination of local spread and transport of cases between cities or regions. Proactively limiting inter-regional mobility as a protective measure can desynchronize regional epidemic waves by delaying the introduction of cases and reducing the number of parallel infection chains \cite{zunker2025risk}. This desynchronization enables the distribution of the public health burden between differently affected neighboring regions, and is thus desirable from a national perspective. 
Hospital beds can be shared across the country only when they are not needed everywhere at the same time.
However, the mismatch between personal experiences and early-introduced interventions may be perceived as unfair and irrational by some individuals, making them particularly susceptible to generating and supporting narratives that reduce their likelihood of complying with restrictions \cite{kominsky2021if,petherick2021worldwide,schumpe2022predictors}. 

Populations that perceive themselves as part of a collective effort are more likely to accept restrictions aimed at limiting disease spread \cite{schumpe2022predictors}. In contrast, when individuals view interventions as disproportionately burdening their group, trust erodes and adherence wanes \cite{bor2023covid,wright2021predictors}. In this sense, the effects of desynchronized waves are not only epidemiological but also social: regions with lower incidence may feel wrongly constrained, precisely when mobility restrictions are most critical for preventing the synchronization of outbreaks.

Preparedness, then, requires a dual strategy. Mechanistic models should be complemented by communication strategies that align narratives of fairness, solidarity, and shared identity with these underlying dynamics. While the mechanical properties of disease spread set the stage, it is narrative and identity that determine how societies navigate these trajectories. By integrating both dimensions, preparedness strategies can achieve both legitimacy and effectiveness.

\begin{figure}[ht!]
    \centering
    \includegraphics{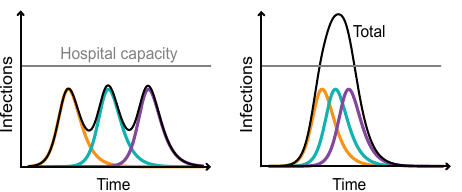}
    \caption{By slowing the spatial spread of epidemics and desynchronizing regional waves (in color), the total number of infections (black) might stay below the hospital capacity limit. However, if these measures appear disconnected from the local reality, people may see them as overreactions, leading to frustration and reduced compliance.}
    \label{fig:desynchronized_waves}
\end{figure}

\ifuseclearpage\clearpage\fi
\section{Solidarity circles change over the course of a pandemic}
\ifuseauthors\textit{Barbara Prainsack and Isabella Radhuber}\fi
\label{mec:solidarity}

The ties that unite societies matter before, during, and after crises. At the same time, crises can also intensify divisions, erode trust, and exacerbate inequalities, with consequences for compliance with interventions and undermining epidemic control  \cite{wef2025globalrisks,radhuber2025inclusive}. It remains debated whether crises foster solidarity or fragmentation \cite{ Mohamed2025WarandSocialSolidarity}. While some studies emphasize the divisive effects of crises, others highlight how acute crises can activate solidaristic mechanisms \cite{HUGGINS2025Place-Based}. In this context, high institutional and interpersonal trust fosters compliance and solidarity, while low trust amplifies perceptions of unfairness, social cohesion, and receptiveness to misinformation. Likewise, perceived risk may initially encourage collective action, but prolonged or unevenly distributed risks can deepen social divisions \cite{devine2021trust, siegrist2021impact}. Solidarity shapes how pandemics unfold by influencing social cohesion, protective behaviors, and compliance with interventions. However, this effect is not necessarily positive, e.g., when populations are divided because some groups face disproportionate impacts, or when in-group solidarity increases hostility towards those outside one’s own group. The boundaries and meanings of solidarity also depend on cultural and political contexts, shaping who is considered part of the moral community and how collective responsibilities are understood and expressed. Therefore, understanding what factors cause solidarity to materialize, grow, and shift is key to preparing for future crises.

Early in the COVID-19 pandemic, many societies experienced so-called inclusive solidarity, cutting across societal divides \cite{kieslich2023solidarity, prainsack_inpress}. This phase was characterized by openness to differences and willingness to support others regardless of their background, social status, or beliefs. Over time, inclusive solidarity often gave way to exclusive (in-group) solidarity---a narrowing of solidaristic commitments to smaller, more homogeneous groups of like-minded individuals \cite{ferwerda2024crises}. Mask-wearers expressed solidarity within their groups, while vaccine skeptics found support in theirs \cite{zimmermann2021face,radhuber2025inclusive}. Digital social environments and virtual communities could both sustain solidaristic ties across physical distance and reinforce segmentation into like-minded groups. For example, online studies reported increased network segmentation and hardened echo chambers \cite{cinelli2021echo, hartmann2025systematic, rabb2023investigating}, with severe consequences for real-world contact networks - but the evidence is still emerging. 
The shift from inclusive to exclusive, particularistic forms of solidarity, is not inevitable but likely when responses ignore the fact that pandemic measures may not impact all groups equally \cite{radhuber2025inclusive}. To prevent such unintended negative consequences, first, narratives that frame crises in terms of deserving and undeserving groups must be avoided. Second, it is essential to establish support systems---social, economic, and psychological---that are effective and inclusive. When people feel abandoned or disproportionately burdened, their capacity for broad-based solidarity declines \cite{kieslich2023solidarity, bulled2023solidarity}.

\begin{figure}[ht!]
    \centering
    \includegraphics{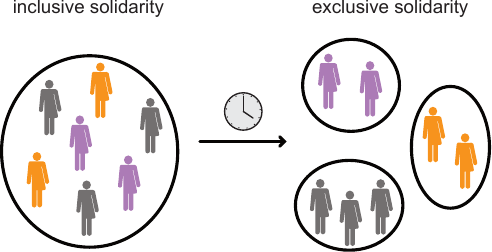}
    \caption{Crises may ignite solidarity and prompt cohesion in societies in distress (left). Yet, prolonged restrictions often lead to fatigue, causing this effect to diminish (right). It is crucial to understand how the cohesive and repulsive forces balance each other and how this balance changes over time.}
    \label{fig:solidarity}
\end{figure}

\ifuseclearpage\clearpage\fi
\section{Feedback between human behavior, spreading dynamics, and policy design}
\label{mec:three_body_problem}
\ifuseauthors\textit{Mirjam Kretzschmar}\fi

Epidemic and pandemic infectious disease outbreaks involve three major players with potentially conflicting aims: the pathogen, which tends to spread within a population; individuals, who aim to avoid becoming sick without restricting their daily activities; and policymakers, who aim to control the epidemic at minimal societal cost. These players interact to form a dynamical system with possibly complex dynamics, meaning that the actions of one player have non-trivial effects on the others. This has far-reaching consequences for how we build and interpret mathematical models for infectious diseases.

A typical example is the adaptation of human behavior in response to current incidence and interventions, which may cause interventions to counteract or complement one another.
For example, the rapid spread of risk awareness and subsequent voluntary adoption of protective behaviors, such as mask wearing and hygiene measures, complemented the mandatory NPIs during the first COVID-19 wave, helping to lower and postpone the epidemic peak~\cite{teslya2020impact}. The same complementarity led to short but strict lockdowns to mitigate more effectively than prolonged moderate measures, as they avoid the effects of pandemic fatigue~\cite{petherick2021worldwide,priesemann2021} and declining adherence \cite{brunekreef2025impact}. 
Conversely, the rollout of COVID-19 vaccinations in 2021 led some vaccinated individuals to increase their contact rates to pre-pandemic levels, partially counteracting the mitigating effect of vaccination and potentially leading to higher short-term incidence \cite{teslya2022importance,bauer2021relaxing,wagner2025societal}. These effects incur complex feedback loops. While we understand their mechanics, their specific dynamics can depend strongly on the societal setting, and are thus hard to predict~\cite{wagner2025societal}.

Modeling the interplay among health opinions that spread through social influence, the pathogens' spread, and policy responses is complex due to the heterogeneous and time-varying nature of these influences. Understanding the combined effect of such mechanisms requires first a deep understanding of each mechanism in isolation. Here, theoreticians need to strike a delicate balance between clarity and simplicity while acknowledging the complexity when translating the principles into real-world settings. A more detailed discussion is provided in Appendix~\ref{SI:three_body_problem}.

\begin{figure}[ht!]
    \centering
    \includegraphics{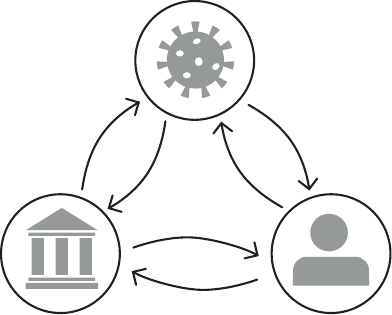}
    \caption{Epidemics operate as a dynamic feedback system combining the competing dynamics of the pathogen, societies, and policymakers. To assess the impact of public health measures and effectively mitigate the disease, none of these three players may be neglected.}
    \label{fig:three_body_problem}
\end{figure}

\ifuseclearpage\clearpage\fi
\section*{Interplay between mechanisms}
\label{sec:interplay}

As discussed in the previous mechanism (\ref{mec:three_body_problem}), the trajectory of a pandemic emerges from the continuous interaction between pathogen characteristics, immunity, human behavior, and policy responses. Consequently, the mechanisms discussed in this work cannot be regarded as isolated principles with universally fixed effects. Instead, their applicability and effectiveness depend on the epidemiological, social, and institutional context in which they operate. Importantly, this context also includes interactions between the mechanisms themselves, which can reinforce, modify, or attenuate one another. 

Several mechanisms describe mathematical or epidemiological relationships under simplifying assumptions, such as homogeneous mixing or stable compliance patterns. However, real-world transmission dynamics are shaped by heterogeneous contact networks, social inequalities, risk perception, and varying levels of institutional trust. For example, behavioral and solidarity-based processes shape compliance with non-pharmaceutical interventions (NPIs), thereby conditioning the effectiveness of mathematically derived mitigation strategies. Social mechanisms, therefore, do not merely accompany epidemiological mechanisms, but often determine not only whether but also how their theoretically predicted effects can emerge in real-world settings. 

The mechanisms compiled in this work should therefore not be viewed as independent, but as interacting. In the following, we discuss several important interactions and limitations that emerge from this interdependence. Importantly, while this section focuses on interactions between mechanisms, our compilation is not exhaustive. Broader societal dimensions---including economic and psychological factors---are likewise part of the same coupled system that shapes pandemic dynamics and decision-making, as discussed below.

\paragraph{Solidarity and the effectiveness of combined NPIs.}
The effectiveness of combined NPIs (\ref{mec:redundant_npis}) depends strongly on the solidarity dynamics (\ref{mec:solidarity}). Mathematical analyses of intervention combinations often implicitly assume stable and homogeneous compliance across the population, overlooking the diverse social and behavioral factors that influence adherence. In practice, however, adherence to policies depends on factors such as trust in institutions, perceived fairness, social support systems, collective identity, and more. Inclusive solidarity can amplify the synergistic effects of combined NPIs by increasing voluntary adherence and mutual reinforcement of protective behaviors. Conversely, a shift in solidarity toward smaller in-groups can weaken compliance. Synergy and redundancy between the same measures can therefore change over time as solidarity circles evolve. This interaction illustrates that the real-world effectiveness of NPIs is not determined solely by their epidemiological properties, but also by the social conditions under which they are implemented.

\paragraph{Airborne transmission changes epidemic dynamics}
The quadratic effect of reductions in gathering sizes on expected new infections (\ref{mec:quadratic_reduction}) depends on the viability of the following assumption: that the infectious potential of an infected attendee is not diluted as gathering size increases. To the extent that a specific pathogen requires prolonged close contact for transmission, this assumption fails since, at a given time, the average distance between participants at small gatherings is less than that at large gatherings. In biological terms, this means the effect relies on the pathogen's capacity for airborne spread, enabling transmission over long distances. Airborne transmission may also be a prerequisite for superspreading of respiratory diseases, since superspreading requires infecting distant or fleeting contacts (\ref{mec:superspreader}). In general, epidemic control and management are a matter of choosing the right measure for the right pathogen, and mechanisms \ref{mec:quadratic_reduction} and \ref{mec:superspreader} may not apply to respiratory pathogens lacking airborne transmission potential.

\paragraph{Superspreading and contact structure.}
Repeated contacts and contacts within clustered groups limit the number of susceptible individuals that can be infected, whereas sporadic non-repeated contacts result in a higher number of overall infections and might connect otherwise separated social clusters, thus, creating opportunities for large transmission chains. Consequently, interventions limiting sporadic non-repeated interactions may disproportionately reduce large-scale spread, as explained in section (\ref{mec:repeated_contacts}). As mentioned there, this is especially relevant for diseases with high transmissibility. Similarly, this reasoning can be applied to superspreading pathogens, which are not necessarily extraordinarily transmissible on average, but where transmissibility is highly heterogeneous and the resulting disease dynamics are dominated by a subset of highly infectious individuals (\ref{mec:superspreader}). For these highly infectious individuals, the restriction of sporadic contacts again provides an opportunity to substantially limit the number of infected individuals, thus reducing the effect of superspreading in the population.

\paragraph{Pandemic establishment and limiting sporadic contacts.}
The usefulness of limiting sporadic contacts (\ref{mec:repeated_contacts}) depends on the probability that an introduced pathogen successfully establishes sustained transmission (\ref{mec:onehealth}). During early outbreak phases, when stochastic extinction remains likely, reducing sporadic transmission opportunities may substantially lower the probability of epidemic establishment, particularly for pathogens with reproduction numbers only moderately above the epidemic threshold. For pathogens with high transmissibility, reducing sporadic contacts alone might not lead to extinction but could substantially slow the spread by limiting offspring to a set of static contact partners.

\paragraph{Solidarity circles and event-size reduction.}
The effectiveness and societal sustainability of event-size reductions (\ref{mec:quadratic_reduction}) depend not only on their epidemiological effects, but also on how fairly such measures are perceived across society (\ref{mec:solidarity}). The quadratic reduction mechanism suggests that moderate reductions across many activities may outperform complete shutdowns of selected sectors. However, implementing broad moderate restrictions requires widespread collective acceptance and trust that burdens are distributed fairly. If some groups perceive themselves as disproportionately constrained while others continue social activities relatively unaffected, solidarity may shift toward narrower in-groups, weakening adherence and increasing social fragmentation. Thus, the practical feasibility of broadly distributed attendance reductions depends strongly on inclusive solidarity and equitable policy design.

\paragraph{Desynchronization, solidarity, and the interpretation of low incidence.}
Another interaction emerges between the desynchronization of epidemic waves across regions (\ref{mec:desynchronized_waves}) and the fact that under successful pandemic management individuals are unlikely to know someone who is ill (\ref{mec:friends_not_sick}). Effective mitigation and mobility restrictions can desynchronize outbreaks across spatially separated populations. This lowers local prevalence at any given time and thereby creates a visibility gap between objective epidemiological risk and individual experience. As a result, precisely those regions where interventions are most effective may experience the weakest subjective signals of risk, reinforcing the prevention paradox described above. Consequently, interventions may appear exaggerated or irrational from an individual perspective precisely because they are effective at the population level. Thus, desynchronization shapes not only epidemiological burden but also public perceptions of risk and necessity.

A second important interaction arises between this visibility gap and the formation of solidarity circles (\ref{mec:solidarity}). When individuals do not personally observe severe illness within their social networks, the interpretation of this absence depends strongly on prevailing social cohesion, trust, and perceived fairness. In contexts of inclusive solidarity, the lack of visible cases can be understood as a collective success of coordinated action, reinforcing compliance and trust in institutions. By contrast, when solidarity fragments into more homogeneous in-groups, the same absence of illness may be interpreted differently across groups, weakening shared narratives about the necessity and fairness of interventions. Consequently, populations exposed to similar epidemiological conditions may nonetheless develop conflicting interpretations of risk and policy legitimacy, amplifying polarization. 

Desynchronization of waves across regions can further strain solidarity, as geographically uneven burdens may be perceived as unjust, even when they reflect effective coordination at a larger spatial scale. In particular, populations experiencing relatively low case numbers may face greater difficulty relating restrictive measures to their own lived experience when the primary benefits lie in preventing outbreaks elsewhere, or averting future transmission. This mismatch between local experience and system-level necessity can undermine perceptions of fairness and weaken solidarity.
Thus, solidarity structures critically mediate whether the invisibility of disease during successful control is interpreted as evidence of effectiveness or as evidence of overreaction, thereby shaping compliance and ultimately influencing epidemic dynamics. Consequently, interventions optimized at the population level may still fail if they are not accompanied by social cohesion, inclusive political measures, and communication strategies capable of explaining their collective rationale.

\paragraph{The effect of exponential growth.}
The exponential growth dynamics in the early phase of pandemics (\ref{mec:exponential_growth}) play a cross-cutting role in several mechanisms discussed in this work. For policy design, exponential growth makes it particularly important to account for the redundancy of measures (\ref{mec:redundant_npis}) and behavioral adaptations that influence their effectiveness (\ref{mec:three_body_problem}). If combined interventions are evaluated under the assumption of additive effects, redundancy may result in insufficient mitigation, allowing continued exponential growth in case numbers. Similarly, behavioral responses can either amplify or weaken intervention effects, where even small changes in effective transmission rates can have large consequences due to exponential amplification. Exponential growth, therefore, reinforces the importance of accurately accounting for effects between and on measures. 
Beyond intervention stringency, exponential growth also highlights the importance of timing in policy responses. Small delays in implementing effective measures, driven by societal responses, can translate into disproportionately large increases in case numbers.
In the case of desynchronization of epidemic waves~(\ref{mec:desynchronized_waves}), early exponential growth implies that limiting spatial spread to enable resource sharing is particularly important during the early phase of a pandemic, before outbreaks across regions become synchronized. 
For One Health-driven establishment dynamics (\ref{mec:onehealth}), exponential growth is embedded in the underlying transmission processes and interacts with stochastic effects and network structure to determine whether early introductions fade out or develop into sustained outbreaks. 
Across these mechanisms, exponential growth links early outbreak dynamics to the timing, coordination, and effectiveness of interventions.
Lastly, the early exponential growth phase of pandemics creates a fundamental challenge for collective action because infection numbers initially appear small despite rapidly increasing transmission potential. Effective early interventions, therefore, require populations to accept preventive measures before the threat becomes directly visible in everyday life. Inclusive solidarity can facilitate such early action by strengthening willingness to adopt protective behaviors for the benefit of others, including vulnerable groups not yet personally affected (\ref{mec:solidarity}). Conversely, if solidarity is weak or fragmented, delayed responses become more likely, allowing exponential growth to continue until the epidemic burden becomes immediately visible. Social cohesion, therefore, influences not only compliance with interventions but also the timing at which societies are willing to act.

\paragraph{Economic and psychological factors.}
Finally, it is important to emphasize that the interactions discussed above are not exhaustive. In addition to interactions between mechanisms, economic and psychological factors are not external “side effects” of interventions, but integral components of pandemic dynamics. Epidemics and mitigation policies jointly shape labor supply, mobility, mental health, and societal behavior, and these effects feed back into transmission and policy effectiveness \cite{eichenbaum2021macroeconomics}. Mitigation measures do entail direct and indirect costs; however, these must be understood in relation to the counterfactual scenario of uncontrolled transmission, which can generate substantial economic disruption and psychological burden in addition to direct treatment-related costs. This trade-off is also reflected in optimal control analyses, which balance intervention costs with infection-driven societal costs. They suggest that, in stylized settings, suppression strategies are optimal for severe pathogens \cite{muller2025optimizing}. While such results depend on strong simplifying assumptions and are not intended as empirical prescriptions, they illustrate the broader principle that the cost of inaction may exceed the cost of intervention, and that economic and psychological impacts are inherently intertwined with transmission dynamics rather than external to them. Nonetheless, understanding  these basic mechanisms can help understand the effect of interventions.

\ifuseclearpage\clearpage\fi
\section*{Outlook}
\ifuseauthors\textit{Carolina J. Klett-Tammen, Manuela Harries, and Philip Bechtle, Seba Contreras, Viola Priesemann}\fi

As witnessed during COVID-19 and previous pandemics, crises often catalyze a surge in theoretical knowledge. However, to render this knowledge actionable, we need to translate abstract theory---often accessible only to specialists---into concrete insights for decision-making. Consequently, our aim was to compile basic mechanisms of pandemic spread and mitigation. These mechanisms are grounded in natural laws and the structure of our societies, governing the emergence, unfolding, and control of pandemics. Understanding them is a prerequisite for policy design.

To identify such mechanisms, we adopted a multidisciplinary approach, as the governing laws emerge at the interface of infectious disease epidemiology and complexity science. These mechanisms are fundamental and, therefore, often not explicitly formulated in study abstracts; moreover, they rarely act in isolation but rather jointly with other effects. Given that this level of abstraction is often considered tacit knowledge within specific fields or is not typically distilled in standard literature, a classical ``systematic literature review'' would fail to extract it. Hence, our work presents the cumulative experience of a diverse group of experts. This compilation is not exhaustive, but it is intended to grow and to inspire abstraction in how we study pandemics, thereby feeding into the broader discussion on pandemic preparedness.

We can only speculate about when and what kind of pandemic threat will emerge next. However, based on the mechanisms we discuss, specific pillars can support preparedness. First, we can only produce meaningful models and scenarios if we have access to proper data, and the earlier, the better \cite{vespignani2020modelling}. This can be achieved by maintaining and adapting robust surveillance and monitoring infrastructure, together with clear rules for distributed data acquisition, access, and use. Second, in the event of a crisis, novel scientific collaboration across disciplines, institutions, and society becomes necessary \cite{rivers2019using}. They are facilitated by Open Source platforms and software, rigorous and timely peer review, and FAIR (Findable, Accessible, Interoperable, and Reusable) data principles~\cite{FAIRprinciples}. Third, strong and independent fundamental research is an immediate asset for pandemic preparedness. We do not know what challenges the next pandemic will bring, but if we have robust, independent fundamental research and a flexible funding and evaluation framework during pandemics, scientists will meet these challenges and work diligently for the benefit of society. 

Compiling these mechanics required contributions across multiple disciplines; mitigating pandemics requires the joint effort of even more experts. One method to coordinate and channel scientific expertise into actionable evidence is to form expert groups. The benefits are clear: when functioning well, such groups provide interactive, transdisciplinary peer review, guide research priorities, and partially protect individual scientists from targeted public attacks. An expert group can compile state-of-the-art evidence alongside quantified uncertainty and distinguish it from unrealistic assumptions across disciplines. However, fundamental questions regarding their governance remain: How are these groups best constituted? Should they be appointed by governments, nominated by independent scientific societies, or self-organize? What is their role in the scientific, political, and public debate? Crucially, the role of science is not to make strong recommendations, but to equip everyone with the best of our current knowledge, both certainties and uncertainties, so that society and politics can jointly reach informed decisions.

Finally, although we have dissected mechanisms in isolation to identify general principles, their combined effects in the real world are as complex as the pandemics they drive. Communicating this non-linear complexity---and the validity and limits of tools such as scenario-based modeling---is challenging by definition. We need to acknowledge the inherent difficulty of this task, particularly amid the current ``infodemic'' of misinformation~\cite{lazer2018science}, and never forget that defeating a pandemic requires joint, long-term efforts from across societies.

\ifuseclearpage\clearpage\fi
\section*{Authors' contributions}
Conceptualization: VP

Methodology: SC, VP 
Validation: all

Writing - Original Draft: all, for details see Appendix~\ref{SI:contributions}

Writing - Review \& Editing: all 
Visualization: LM 
Supervision: SC, VP

Project administration: PD

\ifuseclearpage\clearpage\fi
\section*{Acknowledgements}

VP was supported by the Deutsche Forschungsgemeinschaft (DFG, German Research Foundation) under Germany’s Excellence Strategy - EXC 2067/1-390729940  (MBExC), and the Ministry of Science and Culture of Lower Saxony through funds from the program zukunft.niedersachsen, and is associated with 'CAIMed – Lower Saxony Center for Artificial Intelligence and Causal Methods in Medicine' project (grant no. ZN4257). 

The work of KN and SP is supported by the Federal Ministry of Research, Technology and Space of Germany (BMFTR, grant no. 031L0324B) and TU Berlin. 

The work of UH, LS, KS, and BFN was funded by the Danish National Research Foundation through the PandemiX Center (grant no. DNRF170).

BFN acknowledges funding from the Carlsberg Foundation (grants CF23-0173 and CF24-1337).

The work of PB is supported by the Deutsche Forschungsgemeinschaft (DFG, German Research Foundation) --- project number 460248186 (PUNCH4NFDI).



MA was supported by the Spanish Ministry of Science, Innovation and Universities through the BCAM Severo Ochoa accreditation (CEX2021-001142-S/MICIN/AEI/10.13039/501100011033), and by the Basque Government through the ``Mathematical Modeling Applied to Health'' project and the BERC 2022–2025 program. 
MA  acknowledges financial support from the Spanish Ministry of Science and Innovation (MICINN) through the Ramon y Cajal grant RYC2021-031380-I.

RJNK and MH were funded by the European Union. Views and opinions expressed are however those of the authors only and do not necessarily reflect those of the European Union or the granting authority European Union’s Horizon Europe research and innovation programme. Neither the European Union nor the granting authority can be held responsible for them. The PCR-4-ALL and SUNRISE projects have received funding under the Horizon Europe research and innovation programme (grant agreement No 101095606 and 101073821). For CJK-T and MH, additional funding has been provided by Initiative and Networking Fund of the Helmholtz Association of German Research Centres (grant numbers KA1-Co-08 and SO-096) as well as by intramural HZI funds which supported this work.

The work of AK, VKJ and JS is supported by the Federal Ministry of Education and Research (BMBF)/Federal Ministry of Research, Technology and Space (BMFTR) via OptimAgent (funding number: 031L0299J), RESPINOW (funding number: 031L0298F), PREPARED (funding number: 100665502), NUM-SAR (funding number: 01KX2521), by the German Research Foundation (DFG) via EpiAdaptDiag (funding number: KA 5361/1-1), by the Volkswagen Stiftung via the INFRALINK project, and by the Innovative Medical Research at the University of Münster (IMF) via GetCoSy (funding number: JÄ 122318).

MEK gratefully acknowledges funding from The Netherlands Organisation for Health Research and Development (ZonMw), via project ``NCOH Pandemic Preparedness Research Kickstarter'' (grant number 10710022210003) and project ``Control of Infectious Diseases by Linking Individual and Collective Behaviour in Social Network'' (grant number 10710062310022).


Gemini AI and Grammarly PRO AI were used for grammar checks in the main text. The authors assume full responsibility for the final content of the article.

\clearpage

\end{bibunit}

\clearpage
\begin{bibunit}
\appendix
\section*{Appendix}

\section{Overview of contributions}
\label{SI:contributions}

All authors read and revised all sections. Original drafts of the sections were written by the following scientists:

\begin{itemize}
\item \textbf{Introduction:}\\Seba Contreras, Robyn J. N. Kettlitz, and Viola Priesemann
\item \textbf{Effective prevention requires monitoring the system rather than the outbreak:}\\ Maíra Aguiar
\item \textbf{The effect of combined NPIs can be redundant:} \\ Peter Klimek
\item \textbf{Reducing event size can reduce new infections quadratically:} \\ Kai Nagel and Sydney Paltra
\item \textbf{Limiting sporadic contacts prevents infections between clusters:}\\ Veronika K. Jaeger, Janik Suer, and André Karch
\item \textbf{Early and decisive action limits exponential pandemic growthh:}\\Piklu Mallick, Laura M\"uller, Seba Contreras, and Viola Priesemann
\item \textbf{Superspreading diseases are highly susceptible to restrictions on non-repeated contacts:}\\ Ulrik Hvid, Lone Simonsen, Kim Sneppen, and Bjarke Frost Nielsen
\item \textbf{Your friends are not ill if pandemic control is effective:}\\ Viola Priesemann, Ronja Gronemeyer, Philipp Dönges, and Piklu Mallick
\item \textbf{Desynchronization of waves enables resource sharing:}\\ André Calero Valdez
\item \textbf{Solidarity circles change over the course of a pandemic:}\\ Barbara Prainsack and Isabella Radhuber
\item \textbf{Feedback between human behavior, spreading dynamics, and policy design:}\\ Mirjam Kretzschmar
\item \textbf{Outlook:}\\  Carolina J. Klett-Tammen, Manuela Harries, Philip Bechtle, Seba Contreras, and Viola Priesemann
\end{itemize}

\clearpage
\section{System-level monitoring and the limits of outbreak-based prevention}
\label{SI:system_monitoring}

This section expands mechanism~\ref{mec:onehealth} by clarifying why prevention cannot rely only on detecting outbreaks after human cases have already appeared. The mechanism is based on a mismatch between the timescale on which emergence risk accumulates and the timescale on which outbreaks are usually detected. Spillover and early transmission may be preceded by changes in ecological, epidemiological, behavioural, or environmental conditions, but public-health systems often respond only once human cases are observed and reported. At that stage, the system may already have moved from a low-risk configuration into a high-risk state in which sustained transmission, stochastic amplification, or rapid exponential growth becomes more likely.

This mechanism connects directly to the early-growth mechanism described in Section~\ref{SI:exponential_growth}. Once sustained human-to-human transmission is established, even short delays in detection or response can produce large differences in cumulative incidence. It also interacts with stochasticity and superspreading: near the epidemic threshold, small transmission clusters may either disappear by chance or expand rapidly before routine case-based surveillance detects a clear signal. Thus, prevention based only on observed outbreaks responds to a late manifestation of a broader system-level transition.

The practical implication is that prevention requires monitoring the broader system in which emergence becomes possible. In high-resource settings, this may include wastewater surveillance, genomic monitoring, sentinel surveillance in livestock and wildlife, vector surveillance, environmental monitoring, and integrated data platforms. In lower-resource settings, the same principle may be implemented through community-based surveillance, syndromic reporting, animal-health networks, targeted ecological monitoring, and strengthened communication between human-health, veterinary, and environmental sectors. The mechanism is therefore not tied to a single technology, but to the integration of multiple weak signals across the One Health system.

A key limitation is that upstream warning signals are often noisy, incomplete, and probabilistic. Detecting a change in system risk does not automatically identify which intervention is justified, when it should be triggered, or how costly false alarms may be. This creates a communication challenge: authorities must be able to explain why action may be needed before a visible outbreak occurs. If early warnings are poorly communicated, repeated false alarms may undermine trust and reduce compliance. Thus, the effectiveness of this mechanism depends not only on surveillance capacity, but also on data integration, institutional coordination, public trust, and transparent risk communication.

\clearpage

\section{Two-intervention algebra with absolute overlap}
\label{SI:redundant_npis}

This section provides an extended mathematical framework for mechanism \ref{mec:redundant_npis}.

We consider an SIR-like model with baseline transmission rate $\beta_0$ (no NPIs) and recovery rate $\gamma$. 
Let two NPIs be indexed by $1,2$. Each NPI, when applied alone, removes a fraction $\varepsilon\in[0,1]$ of risky contacts.
Define \(\phi\in[0,1]\) as the \emph{absolute} fraction of contacts removed by \emph{both} NPIs, i.e.\ the measure of the intersection:
\[
\phi = |A\cap B|,
\]
where $|A|=|B|=\varepsilon$ are the (fractional) contact sets removed by NPI~1 and NPI~2 respectively.

\paragraph{Union (combined) reduction.}
The total fraction of contacts removed by the union of the two NPIs equals
\begin{equation}
r_{\mathrm{union}} \;=\; |A\cup B| \;=\; |A|+|B|-|A\cap B|
= 2\varepsilon - \phi.
\label{eq:r_union_abs}
\end{equation}
The transmission rate under both NPIs is therefore
\begin{equation}
\beta \;=\; \beta_0\bigl[1-r_{\mathrm{union}}\bigr]
       \;=\; \beta_0\bigl[1-2\varepsilon+\phi\bigr].
\label{eq:beta_union_abs}
\end{equation}
Consequently the instantaneous effective reproduction number is
\begin{equation}
\mathcal{R}_{\mathrm{eff}}(t)
= \frac{\beta(t)}{\gamma}\frac{S(t)}{N}
= \mathcal{R}_0\frac{S(t)}{N}\bigl[1-2\varepsilon+\phi\bigr],
\label{eq:Reff_union_abs}
\end{equation}
with $\mathcal{R}_0=\beta_0/\gamma$.

\paragraph{Relation to multiplicative (independence) model.}
The standard multiplicative (probabilistic independence) model yields the joint retention factor $(1-\varepsilon)^2$, hence the combined fractional reduction
\begin{equation}
r_{\mathrm{mult}} \;=\; 1-(1-\varepsilon)^2 \;=\; 2\varepsilon - \varepsilon^2.
\label{eq:r_mult_abs}
\end{equation}
Independence corresponds to the special overlap value \(\phi=\varepsilon^2\), because substituting \(\phi=\varepsilon^2\) into eq. \ref{eq:r_union_abs} recovers eq. \ref{eq:r_mult_abs} (under independence, $\varepsilon^2$ is the probability that a contact is a member of contact sets $A$ and $B$ simultaneously). The difference between the union and multiplicative reductions is
\begin{equation}
r_{\mathrm{union}}-r_{\mathrm{mult}} \;=\; \varepsilon^2 - \phi.
\label{eq:diff_union_mult}
\end{equation}
Thus
\begin{itemize}
  \item $\phi>\varepsilon^2$ implies $r_{\mathrm{union}}<r_{\mathrm{mult}}$ (greater overlap than independence; \emph{redundancy}),
  \item $\phi=\varepsilon^2$ implies multiplicative/independent behaviour,
  \item $\phi<\varepsilon^2$ implies $r_{\mathrm{union}}>r_{\mathrm{mult}}$ (less overlap than independence; \emph{synergy}).
\end{itemize}

Note that for the intersection it must hold that $0 \le \phi \le \varepsilon$.
Moreover the union cannot exceed the whole contact set, so \(r_{\mathrm{union}}\le 1\), which implies $\phi \ge 2\varepsilon - 1$.

One may also choose to reparameterise departures from independence by the relative deviation, 
\[
\varphi \;=\; \frac{\phi-\varepsilon^2}{\varepsilon^2},
\]
so that $\varphi=0$ is independence, $\varphi>0$ is excess overlap (sub-additivity) and $\varphi<0$ is deficit overlap (super-additivity). Estimation of \(\phi\) requires data with variation in joint policy deployment; co-occurrence of NPIs limits identifiability and motivates use of staggered rollouts or instrumental variation.

\clearpage
\section {Practical challenges in evaluating NPI interactions}
\label{SI:challenges_redundant_npis}

The theoretical results on NPI redundancy and synergy outlined in mechanism \ref{mec:redundant_npis} have important implications for the design, synthesis, and interpretation of empirical research on intervention effectiveness. While the algebra of NPI interactions can be formalized cleanly in a mathematical framework, the practical evaluation of the effectiveness of combined interventions is challenging and raises several issues, which we address below.

An immediate practical consequence is that effectiveness estimates for individual NPIs cannot be directly compared across studies that differ in the set of co-deployed measures. Because the marginal effect of any single NPI depends on the existing portfolio of controls \cite{haug2020ranking, brauner2021inferring}, estimates of the effect of a given intervention, such as mask mandates, will generally differ depending on which other measures are simultaneously in place, for example, school closures. Direct comparability would require all studies to consider the same set of NPIs being studied in conjunction, a condition that is hardly ever satisfied in practice, given the heterogeneity of pandemic policy responses across countries and time periods \cite{hale2021}. Researchers synthesizing evidence on NPIs drawn from different studies must therefore explicitly account for this co-intervention dependence, and accurately and proactively communicate potential uncertainties that arise from aggregating estimates across heterogeneous policy environments. This comparability problem reflects the deeper structural issue that NPIs do not have fixed, context-independent effects---a point that directly motivates the need for careful study and model design.

Addressing the comparability problem requires moving away from the treatment of NPIs as independent clinical interventions whose effects can be simply added or pooled. Rather, they are highly endogenous, deeply correlated disruptions to a complex adaptive social system \cite{snoeijer2021measuring, thomson2022international, andrews2016impacts,chin2021effect}. At the model-design stage, defaulting to a simple log-linear reduction on transmission of the form $R_t = R_0 - \sum_i \alpha_i X_i$, implicitly assumes independence between measures and will misrepresent the redundancy structure described in Appendix \ref{SI:redundant_npis}. More appropriate functional forms should allow for multiplicative interactions, ceiling effects, and regularized interaction terms that reflect diminishing returns as the portfolio of active measures grows \cite{brauner2021inferring, haug2020ranking}. Including high-frequency behavioral proxies alongside NPI indicator variables---such as aggregate mobile phone mobility data, consumer transaction indicators, or public transport usage---can help to assess behavioral responses \cite{nouvellet2021reduction, persson2021monitoring}. At the cross-study synthesis stage, NPI effect sizes should never be pooled across studies without adjusting for the co-intervention background, for instance, as provided by government response trackers in the case of COVID-19 \cite{hale2021}. In meta-analyses, stratification or weighting should account for the strength of the causal identification strategy, not merely statistical sample size. Even with these refinements in place, however, a more fundamental modeling challenge remains.

At a deeper level, the challenge epidemiological modelers face here is a version of the Lucas Critique from macroeconomics \cite{lucas1976econometric}, which states that historical behavioral relationships cannot be reliably used to predict the effects of a policy intervention because policy changes themselves change behavior. While traditional compartmental models are well-suited to represent biological transmission mechanisms, their parameters are typically estimated under behavioral regimes that may not persist under policy innovations or voluntary behavioral adaptation \cite{funk2010modelling, bedson2021review}. As a result, a model calibrated during a period of voluntary distancing may misrepresent the counterfactual scenario with no formal NPIs, since the behavioral responses to perceived risk are themselves endogenous to the policy environment---the very feedback mechanism discussed in mechanism~\ref{fig:three_body_problem}. To address this in practice, strategies from economics to identify causal effects and avoid structural biases offer increasingly promising future research avenues. These include structural equation modeling to isolate unexpected policy shocks via impulse response functions; integration of behavioral and game-theoretic feedback loops directly into transmission models \cite{bauch2004vaccination, wang2016statistical}; and the application of the causal inference toolkit from econometrics, such as staggered difference-in-differences designs \cite{callaway2021difference}, and regression discontinuity; and synthetic control methods \cite{abadie2010synthetic}. Taken together, these approaches would allow researchers to better separate the causal effect of formal interventions from the confounding influence of endogenous behavioral change, ultimately producing more robust and policy-relevant estimates of NPI effectiveness---and bringing empirical practice closer to the theoretical framework laid out in 
Appendix~\ref{SI:redundant_npis}.

\clearpage
\section{The algebra of the quadratic effect in the SIR model}
\label{SI:quadratic_reduction}

This section provides an extended mathematical framework for mechanism \ref{mec:quadratic_reduction}.

In the classical SIR model, the number of new infections $\Delta I_+$ is proportional to the positive term in the derivative of the infectious compartment ($\beta IS$). Explicitly, if $S$ denotes the number of susceptible, $I$ the number of infectious individuals, and $\beta$ the transmission rate, the dynamics of the infectious compartment are given by
\begin{align}
\dot I = \beta S I - \gamma I,
\end{align}
where $\gamma$ is the recovery rate. The number of new infections is therefore proportional to the product of the number of susceptible and the number of infectious individuals.

Reducing the attendance of an event to a fraction $\kappa$ of the original number of participants reduces on average both the number of susceptibles and the number of infectious individuals present at the event. Mathematically, the new number of infections is reduced quadratically 
\begin{align}
\Delta I_+(\kappa) \sim \left(\kappa I\right)\cdot\left(\kappa S\right) = \kappa^2 I S.
\end{align}

If we consider $N$ events that all have the same fractions of infectious and susceptible people, reducing the attendance in each event to a fraction $\kappa$ gives again the quadratic reduction of new cases 
\begin{align}
\begin{split}
\Delta I_+(\kappa) &\sim \sum_{\text{event}=1}^{N} (\kappa I_{\text{event}})(\kappa S_{\text{event}}) \\
&= \kappa^2 \sum_{\text{event}=1}^{N} I_{\text{event}} S_{\text{event}} \sim \kappa^2 \Delta I_+.
\end{split}
\end{align}

In contrast, reducing the number of events to $\kappa N$ while keeping full attendance at each event yields only a linear reduction
\begin{align}
\begin{split}
    \Delta I_+(\kappa) &\sim \sum_{\text{event}=1}^{\kappa N} I_{\text{event}} S_{\text{event}} \\
    &= \kappa \sum_{\text{event}=1}^{N} I_{\text{event}} S_{\text{event}} \sim \kappa \Delta I_+,
\end{split}
\end{align}
which is less effective to control and eradicate the disease. Note that the original assumption that new cases are proportional to $S \cdot I$ relies on the idea that every individual interacts equally with all others. This assumption breaks down for very large event sizes, where individuals will not interact with more people if attendance increases.

\clearpage
\section{Exponential growth in the early pandemic: the fully susceptible limit}
\label{SI:exponential_growth}

This section provides an extended mathematical framework for mechanism \ref{mec:exponential_growth}. We present two complementary derivations of exponential growth: first, through the branching-process and generation-interval formalism, which yields the discrete-generation intuition used in the main text, and second, through the continuous SIR model, which provides an explicit exponential growth rate.

\paragraph{Exponential growth through branching and generation intervals.} At the onset of a pandemic caused by a novel pathogen, the number of infected individuals is small relative to the population, so virtually every contact is with a susceptible individual. Under this condition, the spread of infection can be approximated by a Galton--Watson branching process: each infected individual independently gives rise to a random number of secondary infections with mean $R_0$ before recovering \cite{watson1875probability}.

Let $I_k$ denote the expected number of infected individuals in generation $k$. By the branching-process recursion,
\begin{equation}
    I_{k+1} = R_0 \cdot I_k,
\end{equation}
which gives the closed-form solution
\begin{equation}
    I_k = I_0 \cdot R_0^k,
    \label{eq:Ik_dicrete}
\end{equation}
where $I_0 = I(0)$ is the  initial number of infectious individuals. This is the fundamental discrete-generation exponential: the number of infections multiplies by $R_0$ with each successive generation interval $T_g$. If $R_0 > 1$, case counts grow; if $R_0 < 1$, the outbreak dies out.

The main-text statement that an intervention introduced $k$ generations later must contain an outbreak $R_0^k$ times larger follows directly from this formula. Table~\ref{tab:r0_growth} illustrates the cumulative multiplicative burden $R_0^k$ for representative values of $R_0$ and $k$, showing how quickly the number of infected individuals increases even for modest reproduction numbers.

\renewcommand{\thetable}{F\arabic{table}}
\setcounter{table}{0}
\begin{table}[h]
\centering
\caption{Multiplicative increase in case burden at the moment of intervention as a function of delay (in discrete generations) and $R_0$}
\begin{tabular}{c l l l}
\hline
$R_0$ &
1 generation late $(\times R_0^1)$ &
2 generations late $(\times R_0^2)$ &
3 generations late $(\times R_0^3)$ \\
\hline
1.5 & $\times 1.5$ & $\times 2.3$ & $\times 3.4$ \\
2.0 & $\times 2$   & $\times 4$   & $\times 8$   \\
2.5 & $\times 2.5$ & $\times 6.3$ & $\times 15.6$ \\
3.0 & $\times 3$   & $\times 9$   & $\times 27$  \\
\hline
\end{tabular}
\label{tab:r0_growth}
\end{table}

\paragraph{Exponential growth through the SIR model.} The branching-process argument above is formulated in discrete generations rather than continuous time. To obtain an explicit growth rate in continuous time and to connect $R_0^k$ to underlying epidemic parameters, we turn to the standard SIR model. We consider the standard SIR model with transmission rate $\beta$, recovery rate $\gamma$, and a population of size $N$ partitioned into susceptible ($S$), infectious ($I$), and recovered ($R$) compartments:\\

\begin{equation}
    \dot S = - \frac{\beta}{N} S I,\quad \dot I = \frac{\beta}{N} S I - \gamma I,\quad \dot R = \gamma I.
\end{equation}

The basic reproduction number is $R_0 = \beta/\gamma$, and the effective reproduction number at time $t$ is $R_{\text{eff}}(t) = R_0 \cdot S(t)/N$.

At the onset of a pandemic caused by a novel pathogen, $S(0) \approx N$, so $S(t)/N \approx 1$ for as long as the cumulative number of infections remains a negligible fraction of $N$. Under this approximation, the equation for $I$ decouples from $S$:\\
\begin{equation}
    \dot I \approx \beta I - \gamma I = \gamma (R_0 -1) I.
\end{equation}

This is a linear autonomous ODE with solution\\
\begin{equation}
    I(t) = I_0 e^{\gamma (R_0 -1) t},
\end{equation}
where $I_0 = I(0)$ is the  initial number of infectious individuals. Within the fully susceptible approximation, evaluating this over $k$ generation intervals $(\Delta t = k/\gamma)$ gives $I(t + k/\gamma) = I(t) \cdot e^{k(R_0-1)}$, which mirrors Eq.~\ref{eq:Ik_dicrete} in the limit $R_0 \to 1$, since $e^{k(R_0-1)} = R_0^k$ requires $\ln R_0 = R_0 - 1$, which holds only to first order in $(R_0 - 1)$.

Now, consider an intervention introduced at time $t^*$ that modifies the transmission rate to $\beta' = \beta (1 - \epsilon)$, reducing contacts or transmission probability by a fraction of $\epsilon$. This gives us a modified reproduction number $R'_0 = R_0 (1 - \epsilon)$, again asssuming $S(t)/N \approx 1$. Thus, the dynamics of $I$ become\\
\begin{equation}
    \dot I \approx \gamma \left(R_0(1 - \epsilon) - 1\right) I.
\end{equation}
This expression is negative or zero, meaning $I$ decreases, if and only if\\
\begin{equation}
    \epsilon \geq 1-\frac{1}{R_0}.
\end{equation}
This gives the intervention stringency needed to avoid exponential growth. Interventions that fall short of this threshold produce $R'_0 > 1$ and merely slow, rather than halt, exponential growth.

Furthermore, combining intervention timing and strength, the total cumulative infections avoided by an intervention of strength $\epsilon$ introduced at time $t^*$, relative to the same intervention introduced later at $t^*+\Delta t$, scales as
\begin{equation}
\Delta_{\mathrm{avoided}}
=
I_0
\left(
e^{\gamma(R_0-1)(t^*+\Delta t)}
-
e^{\gamma(R_0-1)t^*}
\right).
\end{equation}

Factoring out the epidemic burden at time $t^*$ gives
\begin{equation}
\Delta_{\mathrm{avoided}}
=
I_0
\underbrace{e^{\gamma(R_0-1)t^*}}_{\text{case burden at time } t^*}
\left(
\underbrace{e^{\gamma(R_0-1)\Delta t}-1}_{\text{depends only on } \Delta t}
\right).
\end{equation}

The first factor, $e^{\gamma(R_0-1)t^*}$, represents the epidemic burden accumulated by time $t^*$, while the second factor depends only on the delay $\Delta t$. For fixed $\Delta t$, this second factor is constant, implying that $\Delta_{\mathrm{avoided}}$ grows exponentially with $t^*$ at the same rate, $\gamma(R_0-1)$, as the uncontrolled epidemic itself.

Equivalently, delaying an intervention by a fixed amount of time always incurs the same fractional increase in infections, but because the epidemic burden itself grows exponentially, the absolute number of additional infections also grows exponentially with the delay. In this sense, the cost of waiting increases rapidly during the exponential growth phase.

Concretely, delaying the intervention by one generation interval $(\Delta t = 1/\gamma)$ corresponds to an additional fraction
\begin{equation}
1-e^{-(R_0-1)}
\end{equation}
of the cases present at time $t^*$. This fraction is approximately $63\%$ for $R_0=2$ and $86\%$ for $R_0=3$.

The fully susceptible approximation holds well as long as $S(t)/N\approx 1$. For typical early-pandemic parameters, this is satisfied for many generation intervals, which is precisely the window in which policymakers need to act early and act sufficiently strongly, although they do not have much information yet. Once the fraction of infectious individuals becomes non-negligible, $R_{\text{eff}}(t)$ decreases endogenously as susceptibles are depleted, and the approximation breaks down.

\clearpage
\section{The prevention paradox: network structure, information environment, and cultural context}
\label{SI:friends_not_sick}

This section provides an extended discussion for mechanism \ref{mec:friends_not_sick}. We examine how the architecture of modern social networks, the information environment they generate, and the cultural and institutional context in which individuals form beliefs together shape the severity and resolution of the prevention paradox.

\textbf{Why the paradox is structurally inevitable.} The core mismatch between subjective experience and epidemiological risk follows directly from the conditions under which mitigation is necessary. A pathogen with high severity and a finite healthcare system implies a very low prevalence threshold at which the system becomes overwhelmed; to stay below it, interventions must suppress transmission. Under effective control, the probability that any given individual personally encounters an active case in their immediate social circle is, by design, negligible. This absence of visible illness can, in turn, erode support for the very intervention that produced it---a self-undermining dynamic. It can only be resolved through communication that explicitly frames the absence of illness as the signature of a successful intervention rather than as evidence against it \cite{messinger2020paradox}. Institutional trust plays a key role here: where trust in public health authorities is high, populations are more likely to accept this and maintain compliance \cite{bargain2020trust, goldfinch2022cross}.

\textbf{How online networks decouple personal experience from local transmission dynamics.} In an offline social network under effective local control, one's immediate social environment reflects, at least approximately, local transmission dynamics, and the low number of visible cases within it is a meaningful signal about local prevalence \cite{giese2025social}. Modern online networks introduce a structurally distinct information channel that disrupts this correspondence. Content from regions with very different epidemiological situations enters the same feed as content from one's own region, and its composition is governed by recommendation algorithms rather than by geography or social proximity \cite{lerman2016information, santos2021link}; at the same time, dramatic or emotionally engaging cases are systematically amplified relative to typical ones \cite{caldarelli2025physics}. As a result, a successfully controlled local epidemic can register as either invisible or catastrophic depending on the composition of one's feed, with neither impression reliably tied to actual local prevalence. A qualitatively different distortion arises from social ties that exist only online: regular contacts in gaming communities, topic-specific forums, or group messaging platforms are subjectively experienced as part of one's social circle \cite{arroyo2019online, prochnow2020social}, yet transmit almost no embodied signals of illness. Parasocial relationships with high-reach content creators compound this further, as audiences weigh their reported experiences as they would those of an acquaintance, so a single creator's narrative about illness---or its deliberate minimisation---can dominate perceived severity across an entire follower base out of all proportion to what a single individual's experience can tell us about the population \cite{lee2021video, walter2022making, harff2021social, schmuck2024popular}. Both effects sever the link between one's felt social environment and the local transmission dynamics on which the personal-network signal otherwise depends.

\textbf{Cultural and institutional context.} Cross-national analyses suggest that voluntary adherence to protective measures tends to be higher in societies whose cultural orientation places greater weight on collective obligations relative to individual autonomy \cite{dinero2022association, melton2021culture, santana2022lessons}, and that more fluid social structures are associated with faster epidemic growth, partly because effective contact networks are larger and less locally clustered \cite{volz2011effects, miller2009spread, petruzelli2021analyzing, colman2021social}. In such settings, the personal-network signal is a weaker proxy for community prevalence even in the absence of online confounding. This interacts directly with the solidarity dynamics described in mechanism \ref{mec:solidarity}: in contexts of inclusive solidarity, the absence of visible illness can be collectively understood as a sign of shared success, reinforcing compliance; when solidarity fragments into narrower in-groups, the same absence is more readily reinterpreted as evidence that the threat was overstated \cite{borinca2024nudging, tsang2021boundaries, murray2012threat}. Effective pandemic communication must therefore not only address the inferential gap between ``I do not see illness around me'' and ``the intervention is working,'' but do so in a manner calibrated to the network structure, media ecology, and cultural context in which that inference is made.

\clearpage
\section{Extended version of ``Feedback between human behavior, spreading dynamics, and policy design''}
\label{SI:three_body_problem}

This section provides an extended discussion of mechanism \ref{mec:three_body_problem}. Epidemic and pandemic infectious disease outbreaks involve three major players with potentially conflicting aims: the pathogen, which tends to spread within a population; individuals, who aim to avoid becoming sick without restricting their daily activities; and policymakers, who aim to control the epidemic at minimal societal cost. These players interact to form a dynamical system with possibly complex dynamics, meaning that the actions of one player have non-trivial effects on the others. This has far-reaching consequences for how we build and interpret mathematical models for infectious diseases.
A typical example is the adaptation of human behavior in response to current incidence of infection and hospitalizations and implemented interventions, which may cause interventions to counteract or complement one another. This adaptation is guided by people’s risk perception and their preferences regarding health-related behaviour, which can be assessed via surveys or other tools \cite{li2025general,michie2011behaviour}. 

Importantly, the channels through which risk perception is formed have expanded with the growing dominance of social media in every communication and information exchange. Online platforms now rapidly diffuse both accurate health information and misinformation \cite{desai2022misinformation}. Algorithmic curation tends to reinforce pre-existing beliefs and foster echo chambers, in which users are predominantly exposed to like-minded content, making it harder to update risk perceptions in response to new information \cite{cinelli2021echo}. It is increasingly difficult for policy makers to compete with these social media influences in communicating on preventive measures, but social media sentiments have to be taken into account to predict adherence \cite{williams2023effectiveness}. 

On the other hand, behaviour is determined by control measures implemented by health authorities. These can be varying in stringency and adherence to these measures depends on societal and cultural factors \cite{hale2021, brauner2021inferring, haug2020ranking}. For example, the rapid spread of risk awareness and subsequent voluntary adoption of protective behaviors, such as mask wearing and hygiene measures, complemented the mandatory NPIs during the first COVID-19 wave, helping to lower and postpone the epidemic peak \cite{teslya2020impact}. The same complementarity led to short but strict lockdowns to mitigate more effectively than prolonged moderate measures, as they avoid the effects of pandemic fatigue \cite{petherick2021worldwide,priesemann2021} and declining adherence \cite{brunekreef2025impact}. Conversely, the rollout of COVID-19 vaccinations in 2021 led some vaccinated individuals to increase their contact rates to pre-pandemic levels, partially counteracting the mitigating effect of vaccination and potentially leading to higher short-term incidence \cite{teslya2022importance,bauer2021relaxing,wagner2025societal}. A example of the latter is when a short term lifting of social distancing measures for vaccinated individuals (assessed with a digital Covid certificate) led to a large peak in incidence in the Netherlands in July 2021 \cite{van2023sars}. These interactions between changing immune status of the population, people’s risk perception, adherence to measures, and policy decisions about interventions incur complex feedback loops. While we understand their mechanics, their specific dynamics can depend strongly on the societal setting, and are thus hard to predict \cite{wagner2025societal}. These effects add to complexities of infection dynamics that are influenced by the interaction between host and pathogen, in particular by the host’s immune response and the built up and waning of immunity during and after epidemic waves \cite{molla2023pharmaceutical}. Modeling the interplay among health opinions that spread through social influence, the pathogens’ spread, and policy responses is complex due to the heterogeneous and time-varying nature of these influences. Understanding the combined effect of such mechanisms requires first a deep understanding of each mechanism in isolation. Here, theoreticians need to strike a delicate balance between clarity and simplicity while acknowledging the complexity when translating the principles into real-world settings.

\end{bibunit}

\end{document}